# Calculation of The Lifetimes of Thin Stripper Targets Under Bombardment of Intense Pulsed Ions

S.G.Lebedev[1] and A.S.Lebedev[2]

[1]Institute for Nuclear Research of the Russian Academy of Science,
[2]Lomonosov Moscow State University, Faculty of Calculus Mathematics and Cybernetics

**Abstract**
The problems of stripper target behavior in the nonstationary intense particle beams are considered. The historical sketch of studying of radiation damage failure of carbon targets under ion bombardment is presented. The simple model of evaporation of a target by an intensive pulsing beam is supposed. Stripper foils lifetimes in the nonstationary intense particle can be described by two failure mechanisms: radiation damage accumulation and evaporation of target. At the maximal temperatures less than 2500 $^oK$ the radiation damage are dominated; at temperatures above 2500 $^0K$ the mechanism of evaporation of a foil prevails. The proposed approach has been applied to the discription of behaviour of stripper foils in the *BNL* linac and *SNS* conditions.

## 1. Introduction

Thin solid-state targets are widely used in the world for stripper of ions in accelerators of the charged particles. Thickness of stripper targets is defined by type, energy, and also a charge of ions before and after stripping. The processes occurring in a target under ion bombardment, here some decades are a subject of steadfast studying. Especially intensive research of mechanisms of destruction of targets in ion beams was in 70-80-*th* of the XX century - during a rapid development of high intensity accelerators of elementary particles, methods of ionic implantation, creation of installations for acceleration of heavy nuclei. Recently interest to understanding of processes in solid-state targets has renewed on a wave of creation of new high-intensity and high-energy accelerators of the charged particles, such as *Fermilab* and *SNS* in the USA and *J-PARC* in Japan.

From the point of view of efficiency of obtaining of higher charging states of a beam, loading of vacuum system and small overall dimensions using of solid stripper devices is preferable in comparison with gaseous ones.

The basic disadvantage of all solid-state stripper targets is their limited serviceability, which limits an overall performance of accelerators and raises radiating loading of the personnel serving targets.

From the very beginning of use of thin stripper targets and till now there is a struggle for improvement of manufacturing techniques with the purpose of increase in their service life. The greatest distribution as a material for thin solid-state targets was reached with carbon owing to its high melting point and mechanical durability. Because of its small nuclear weight carbon brings the minimal distortions in the parameters of a particle beam. In the end of 70-th of the XX century a perspective method of manufacturing of carbon foils have been developed – the cracking of hydrocarbon in the glow discharge [1] which for the long time was used for manufacturing of the most long-living carbon foils. Later on the method of manufacturing of carbon foils by means of laser ablation [2], which has given new increase in service life, has been developed. Recently the improvement of technology of thin solid-state targets and increasing of their lifetime are take place due to efforts of Isao Sugai (*KEK*) which has developed the whole series of methods of manufacturing of carbon targets. It had been developed methods of the arc category controllable on a direct current (*CDAD*), controllable on constant and to an alternating current of the arc discharge

(*CADAD*), the mixed ionic beam sputtering (*MIBS*), high ion beam sputtering (*HIBS*) [3] and other methods.

**2. Historical sketch of studying of behavior of carbon targets under ion bombardment**

The technique of quantitative estimations of the lifetime of carbon stripper targets was developed from the very beginning of their uses. From behavior of an irradiated foil follows, that the foil lifetime $t$ strongly depends on its crystal lattice destruction, therefore $t$ should be in inverse proportion to speed of displacement of atoms of target $K_d$. For the first time semi empirical formula connecting $t$ and $K_d$, has been offered in [4]. The big income to the development of model of destruction of carbon films in the ion beams has been contributed by J.L.Yntema from Argonne National Laboratories (*USA*) and F.Nickel (*GSI*), which have in details estimated the value of the speed of displacement of atoms of the target $K_d$ [5]. For heavy ions of high energy the speed of displacement can be expressed as follows:

$$K_d = \frac{N_0 \bar{\varphi}}{E_d} \int_{E_d}^{W_{max}} W d\sigma(E,W) \quad , \quad (1)$$

where $N_0$ – is the full number of atoms of a target, $\bar{\varphi}$ - the density of a flux of bombarding particles averaging over a time, particles/cm$^2$/sec, $W$ – the recoil energy of the displaced atom of a target, $d\sigma$ - the differential cross-section of scattering for single collision of an ion with atom of a target,

$$W_{max} = \frac{4 M_1 M_2 E}{(M_1 + M_2)^2} \quad (2)$$

- the maximal energy transferred at such collision, $E_D$ - energy of displacement of atom of a target. The basic assumption consisted that the target collapses if half of its atoms is displaced from initial positions [6]. Then the lifetime due to destruction of a lattice can be estimated so:

$$t_D = \frac{E_d M_2 E}{2\pi \bar{\varphi} M_1 Z_1^2 Z_2^2 e^4} \left[ \ln\left(\frac{W_{max}}{E_d}\right) \right] \quad . \quad (3)$$

Last expression was obtained with the using of Reserford's differential cross section, which is a good approach at the ion specific energy above 0.1 MeV/nucleon. It has appeared, that the equation (3) badly describes the experimental data because of neglecting influence of heating of a target during an irradiation. In the work of J.Yntema [7] it has been shown, that heating of a foil increases its lifetime. This circumstance has been considered in the work [6]. It has appeared possible to describe well the experimental results on the lifetimes of carbon foils by means of one semi empirical formula considering both displacement of atoms of a target and heating:

$$t = t_D a \exp(-\frac{b}{T}) \quad , \quad (4)$$

where $a$ and $b$ - the empirical constants obtained by means of a method of the least squares from the experimental data, $T$ - the average temperature of a target defined by means of Stephen-Boltsman relation:

$$T = \left(\frac{P}{2\varepsilon\sigma} + T_0^4\right)^{\frac{1}{4}}, \quad P = \frac{dE}{dx}kh\overline{\varphi}, \qquad (5)$$

where $dE/dx$ ($MeVcm^2/g$) – the electronic stopping power of the ion in a material of a target, $h$ – the thickness of a target, $g/sm^2$, $k = 1.6 \cdot 10^{-13}$ J/MeV – the transer factor, $T_0$ - an ambient temperature, $\varepsilon$ – the radiating ability of a material of a foil, $\sigma = 5.67 \cdot 10^{-12}$ $W/cm^2K^4$ – the Stephen-Boltsman's constant. The kind of the equation (4) allows drawing an analogy with the annealing of radiation defects with the activation energy $b$. The targets lifetime increase with the increase of temperature, and, means, the thickness proves to be true by the numerous experiments. However in the work [8] the experimental dependence (see Fig.1) of the lifetime reduction with increase of a target thickness is presented. However the inverse relationship is caused by the reduction of transmission of ions at the increase in thickness of a target above equilibrium so the resulted dependences can be considered as a funny thing. The increase in the lifetime of a stripper foil at increase in thickness occurs only at rather low temperatures T <2500°K when there is no evaporation of a material of a foil. In conditions of evaporation as it will be seen below, the lifetime of a target decreases with increase in its thickness. In work [6] it is shown, that for enough thick targets ($h>10$ $\mu g/sm^2$) the contribution of sputtering of atoms in comparison with radiation damage can be neglected.

The stated technique has connected the destruction of carbon targets with the real processes in solid state – the creation of defects under irradiation, their migration and recombination. Lack of the given technique is its empirical character. There is a natural desire to connect the empirical constants in the formula (4) with the real processes in a foil under an irradiation. Though the sense of a constant $b$ is intuitively clear, nevertheless it is necessary to connect its value with real value of the migration energy of the certain kind of point defects in the real nanostructured carbon films. As if to $a$ constant unfortunately it is not possible to relate its sense with the displacement of atoms of a lattice. The question arises: why the failure of a film takes place at the displacement of half of atoms from their initial positions?

### 3. Influence of radiation damage on the foil lifetime

The author of given paper some years has devoted to the research of behaviour of thin film targets under ion bombardment [9-13]. In the work [9] the relation of a kind (4) for the lifetime of a carbon foil has been deduced from the first principles of physics of radiation defects and the stress - deformed state of solid state. It is known, that in the solid state at irradiation the two kinds of point radiation defects are created - the displaced atoms (or interstitials) and vacancies. The displaced atoms possess the high mobility, and the most part from them annihilate with vacancies in a lattice, and the remained form the molecular complexes. The sizes and number of the complexes depend on temperature of an irradiation, speeds of defect generation, etc. the vacancies which have formed at an irradiation usually remain isolated and inactive at low temperatures, but at high enough temperatures also get mobility. As a result of an irradiation there is an accumulation of radiation defects, which deform a crystal lattice. Around of vacancies there is a compression of a lattice. Under mobility of the displaced atoms a part from them recombines with the vacancies, other part creates the complexes. As a result, the big number of vacancies remains isolated and cause all-round compression of a lattice of a crystal. The deformation of a crystal lattice causes internal pressure in the foil. If these pressure reach the ultimate strength of a foil then it destruct. This is the picture of destruction of carbon foils under irradiation, which create the physical basis for empirical expression

(4). Omitting the detailed calculations presented in [9-11], we shall write out resulting relation for the lifetime:

$$t = 0.23 \left( \frac{3\rho_i \sigma_P}{\xi \Delta \rho \, M} \right)^{\frac{3}{2}} \frac{v^{\frac{1}{4}}}{K_d^{\frac{5}{4}}} \exp\left( -\frac{E_m^i}{4 k_B T} \right) . \qquad (6)$$

In the expression (6): $\sigma_P$ - is the ultimate strength, $M$ - the elasticity module, $v$ - the oscillation frequency of atoms in the lattice ($5 \cdot 10^{13}$ Hz), $E_m^i$ – the migration energy of the displaced atoms in the foil, $k_B$ – Boltsman's constant, $\Delta\rho = \rho_i - \rho_f$ – the change of density of a material of a foil due to an irradiation, $\rho_i$ – the density of an initial phase, $\rho_f$ - density of a final phase. The factor $\xi$ defines the conditions of fastening of a film on the film frame (in case of rigid fastenings of a flat foil on the frame $\xi = 1$).

For an estimation of the speed of atom displacement it is possible to use the well-known expression:

$$K_d = \frac{S_n \overline{\varphi}}{2 E_D}, \qquad (7)$$

where $S_n$ - characterizes energy losses of a moving particle on the formation of defects. Expression for $S_n$, describing experimental data on elastic scattering of ions on atoms looks like [14]:

$$S_n(E) = \frac{4\pi a Z_1 Z_2 e^2 M_1}{M_1 + M_2} S_n(\varepsilon), \qquad (8)$$

$$S_n(\varepsilon) = \frac{1.7\sqrt{\varepsilon}(\ln \varepsilon + e)}{1 + 6.8\varepsilon + 3.4\varepsilon^{\frac{3}{2}}}, \qquad (9)$$

$$\varepsilon = \frac{a M_2 E}{Z_1 Z_2 e^2 (M_1 + M_2)}, \qquad (10)$$

$$a = \frac{0.9 a_0}{\left( \sqrt{Z_1} + \sqrt{Z_2} \right)^{\frac{2}{3}}}. \qquad (11)$$

For the carbon $E_D = 25$ eV, $a_0 = 0.53 A$. Expression (9) take into account the shielding of Coulomb's interactions at $\varepsilon \leq 10$, at $\varepsilon \gg 10$ it is possible, by analogy to work [6] to use the Rutherford's expression for differential cross section:

$$S_n = \frac{\ln \varepsilon}{2\varepsilon} . \qquad (12)$$

The migration energy $E_m^i$ of the displaced atoms is connected with the crystallite melting point $T_m$ by means of the relation:

$$E_m^i = k T_m. \qquad (13)$$

As it is known, at the reduction of crystallite size $L_c$ its melting point decreases according to the Thomson's formula:

$$T_m(L) = T_{m0} \exp\left( -\frac{2\sigma_T}{L_\kappa \rho_\kappa \Delta H_0} \right), \qquad (14)$$

where $T_{m0} = 4800\ ^oK$ –the temperature, $\Delta H_0 = 10$ kJ/g –the heat of fusion of infinitely big crystallite [15], $\rho_\kappa = 1.7$ g/cm$^3$ – the average density of structure, $\sigma_T = 5.5 \cdot 10^{-4}$ J/cm$^2$ - the free surface energy of crystallite. The numerical values are presented for the graphite. Then for a case nanocrystalline graphite with the size $L_c=20A$ that is characteristic for carbon foils obtained by cracking of ethilene in the glow discharge (*GD* - foils), the melting point will decrease up to $3473^oK$. Then for *GD* - foils it is had the resulting formula for calculation of service life:

$$t = 50 K_d^{-\frac{5}{4}} \exp\left(-\frac{870}{T}\right). \quad (15)$$

The expressions (6) and (15) for the lifetime of a carbon stripper foil under ion bombardment allow to take into account all important parameters: the temperature, the rate of creation of displacement (or radiation damage) in a foil, the strength characteristics of a foil material, the migration energy of the displaced atoms and its dependence on the crystalline size, the conditions of fastening of a foil on the frame, the oscillation frequency of atoms in a crystal lattice. The account of such wide set of factors in the rated expression for the lifetime opens opportunities of the description of behaviour under an irradiation carbon stripper targets obtained in various technological processes. The problem of calculation is that at an experimental research of behavior of stripper target under ion bombardment, the measurements of the specified parameters, frequently, are not carry out.

Experience of use of the expression (15) for a prediction of behaviour of stripper targets in in the ion beams testifies to the satisfactory description of temperature dependence and a lifetime. As to displacement rate $K_d$, in comparison with semi empirical formulas (3) and (4) in which the lifetime is in the inverse proportion with $K_d$, the expression (15) predicts nonlinear dependence $t\sim(Kd)^{-1.25}$ which would be interesting for checking up experimentally. From expression (6) it is visible, that $t \sim \left(\frac{\sigma_P}{M}\right)^{\frac{3}{2}}$ and it also would need to be checked up in the experiment. The values of strength characteristics of carbon materials have a wide scatter depending on the manufacturing techniques. In the work [16] the amorphous hydrocarbonic *CVD* foils at a density of 2.19 g/cm$^3$ for the elastic module M have the value of 589 GPa that makes about 52 % from the elastic module of the diamond. Hoshino et al. [17] reported that hydrogenated *DLC* films deposited in dc plasma of methane and hydrogen gas mixture at a particular anode and substrate position had a high value of 850 GPa for Young's modulus using this method. This Young's modulus is about 74 % of the directionally averaged value of diamond, 1141GPa [18]. In the work [19] for carbon nanotubes $M=1000$ *Gpa*, ultimate strength for nanotubes and diamond $\sigma_P=100$ GPa, for graphite $M=10$ *GPa*, $\sigma_P=50$ *MPa* has been obtained. Limiting value of ultimate strength for *CVD* foils is $\sigma_P=47$ *MPa* [20]. The ultimate strength for *GD* - foils according to authors of work [21] makes 27 *MPa*. Authors of work [22] for CVD a foil obtain the value $\sigma_P=300$ *MPa*. A film, obtained by the laser plasma ablation cosistinf of 75 % sp$^3$ and 25 % sp$^2$ coordinated atoms has the elastic module $M=369$ *GPa* [23]. The decreasing of quantity of a diamondlike phase gives rise to decrease of the elastic module. The elastic module of foils, obtained by ion beam deposition, containing 0-16 % of sp$^3$-phase, varies within the limits of from 100 up to 260 *GPa* [24]. In all presented experiments it was measured either ultimate strength, or the elastic module, however for the analysis of service life of stripper targets as it follows from expression (6), it is necessary to know simultaneously the both values. From the presented data it is seen, that the strength characteristics of carbon materials strongly depend on manufacturing techniques. For the diamond and nanotubes $\sigma_P \sim 0.1M$, that characterizes limiting value of strength. In expression (15) for *GD* - foils the value $\sigma_P/M=0.01$ was used.

**4. Evaporation of a target by an intensive pulsing beam**

At the designing intensive accelerators of the charged particles, such as *SNS* the estimation of the lifetime of stripper targets is important for definition of efficiency and radiation load on the persopnnel. Modelling of the behaviour stripper targets under an irradiation in *SNS* was carrying out on the *BNL* linac [25]. Parameters of a *H*⁻ beam of *SNS* and modeling beam of in *BNL* are presented in Table 1.

Table 1. The parameters of *H*⁻ beams in *SNS* and *BNL* linac.

|  | Energy | Duration of an impulse | Frequency | The maximal current | The beam size |
|---|---|---|---|---|---|
| *SNS* | 1 GeV | 1 ms | 60 Hz | 32 mA | 3x2 mm² |
| *BNL* linac | 750 keV | 0.5 ms | 6.7 Hz | 2.02/2.2 mA | O3 mm |

At the analysis of behaviour of the *SNS* stripper target except of radiation damage is necessary to consider also the evaporation of a target due to intensive heating by a circulating H⁻ beam. The similar problem was solved in the work [26] for stationary beams.

The increase in energy losses of a bombarding beam in a target due to multiturn injection leads to it's heating up to 2500 – 4800 $^oK$. It causes sublimation of atoms from a surface of a target and its thickness decreases. On the other hand with reduction of thickness beam energy losses in a target decreases. The temperature of a target decreases, process of sublimation is slowing down, and thickness of a target gets new stationary value $h_1 < h_0$. Thus, processes of a heating, sublimation and change of thickness of a target are interdependent, and it must be taken into account at the description of evaporation of a target under an irradiation.

Pressure of saturated vapor of carbon $P_c$ at a surface of a target essentially depends on its temperature. In a range of temperatures 1700 – 5000 $^oK$ this dependence can be presented in the form of:

$$P_c(T) = A \exp\left(-\frac{B}{T}\right), \qquad (16)$$

where $A=1.87 \cdot 10^{11} Topp$, $B=8.35 \cdot 10^4$ $^oK$. The average speed of movement of vapor atoms of carbon can be estimated so:

$$V = 1.5 \cdot 10^4 \sqrt{\frac{T}{M_2}} \; (cm/sec). \qquad (17)$$

The dependence of density of the carbon sutureted vapor $n$ ($cm^3$) near a surface of a target on the vapor pressure is given by the expression:

$$n = 0.996 \cdot 10^{19} \frac{P_c}{T}. \qquad (18)$$

The quantity of atoms $N_S$ containing in 1 $cm^2$ of a surface of a target with the thickness h and density $\rho$, is calculated as:

$$N_S = 6 \cdot 10^{23} \frac{h\rho}{\mu_c}, \qquad (19)$$

where $\mu_c = 12$ g/mole – is the molar weight of carbon. The number of the atoms leaving 1 $cm^2$ of a surface of a target from its both sides equally $nV$. Then for the speed of change of value $N_S$ the equation is fair:

$$\frac{dN_S}{dt} = -nV. \qquad (20)$$

Differentiating on time the equation (19), we obtain:

$$\frac{dN_s}{dt} = 6 \cdot 10^{23} \frac{\rho}{\mu_c} \frac{dh}{dt}. \qquad (21)$$

Substituting (21) in (20), and using (16) - (19) one can obtained the following differential equation which allows to calculate the dependence of thickness of a target $h$ vs time:

$$\frac{dh(t)}{dt} = -8{,}12 \cdot 10^{10} \frac{\exp\left[\frac{-83500}{T}\right]}{\sqrt{T}}. \qquad (22)$$

Unlike the case of the stationary heating considered in the work [26], feature of loading of stripper targets at *SNS*, *BNL* and many other newly installations is pulsing character of a bombarding beam. The average temperature of a target at $h_0$=200 $\mu g/cm^2$ makes only 997°K therefore evaporation is improbable. However calculations [25] show, that at the same thickness the peak temperature in an impulse reaches 2350°K. Therefore at the analysis of experimental data on stability of a target in *BNL* linac conditions it is important to consider pulsing character of $H^-$ beam. The feature is that the evaporation occurs in that time interval of an impulse when the target is bombarded by ions, and the temperature of heating exceeds a threshold of sublimation.

For the adequate description of processes of heating, cooling and evaporation of a target in a pulsing mode similarly the work [27] it is used a non-stationary heat conduction equation, as a result for the description of the interconnected processes of heating and evaporation of a foil in a pulsing beam of ions we can obtain a following set of equations:

$$\frac{dT}{dt} = \frac{1}{c(t)h(t)}\left(P(t) + 2\varepsilon\sigma_0 T_0^4 - 2\varepsilon\sigma_0 T^4(t)\right), \qquad (23)$$

$$\frac{dh(t)}{dt} = -8{,}12 \cdot 10^{10} \frac{\exp\left[\frac{-83500}{T}\right]}{\sqrt{T}}, \qquad (24)$$

where $P(t)$ – is the pulse power, which is defined as:

$$P(t) = \varphi(t) \cdot \frac{dE}{dx} \cdot h_0 \cdot k,$$

$$\frac{dE}{dx} = 275 \frac{MeV \cdot cm^2}{g}, \qquad (25)$$

$\varphi(t)$ – is the pulse density of a flux of the bombarding particles, similar to the value of $\overline{\varphi}$, but defined through a pulse current, $T_0 = 293°K$ - the initial temperature of a film. In the equation (24) the dependence of a thermal capacity $C$ vs temperature in the form of [27] is considered:

$$C(T) = 0.0127 + 2.872 \cdot 10^{-3} T - 1.45 \cdot 10^{-6} T^2 + 3.12 \cdot 10^{-10} T^3 - 2.38 \cdot 10^{-14} T^4. \quad (26)$$

### 5. Results of calculation and discussion

The solution of a set of equations (23) - (24) was made by numerical methods by means of program complex *MATLAB7.0* (The MathWorks, Inc.). On Fig.2 the calculation results of temperature of a *BNL* linac target is presented to the first second of an irradiation at a pulse current of protons 2 *mA*. The maximal temperature makes 2532 °K, that exceeds the value of 2350 °K, obtained for a similar case in [25], distinction can be caused by use of a temperature-dependent thermal capacity (26). The offered algorithm of calculation allows considering really effect of interference of heating and evaporation. On Fig.3 and Fig.4 the temperature and change of thickness of a foil under a pulsing beam of *BNL* linac with a pulse current 3 *mA* are shown. As it is possible to see the speed of evaporation and a temperature field of a foil vary in time that is connected with interference of heating and evaporation. Results of calculation of a lifetime of a target due to evaporation are presented on Fig.5 (please compare with Fig.4 of [25]). At calculation it was supposed, that a lifetime corresponds to reduction of thickness of a foil twice, though the given value demands the further refinement. On conditions of experiment [25] for the lifetime value the time of reduction of a current through a foil on 10 % was accepted. There is a question: to what change of thickness there corresponds the given change of a current? For the answer it is necessary to construct the dependence of change of a current behind a target from change of its thickness. On Fig.5 the results of calculation of the lifetime of a target at the *BNL* linac conditions, caused by the mechanism of radiation damage also are presented. At calculations of temperature on the mechanism of radiation damage the relation (5) for stationary heating were used. It is visible, that at the maximal temperatures less than 2500 $^oK$ the radiation damage are dominated; at temperatures above 2500 $^0K$ the mechanism of evaporation of a foil prevails. As a whole the resulting curve made of two complementary pieces, well describes experimental data of work [25]. As for the SNS stripper foil itself our calculation show that the maximum temperature will be 2650 $^oK$ and the lifetime about 0.5-1 hours, so there is some difference from the modelling case of Fig.5 which can be attributed to difference in the beam time structure. However this result requares some refining of differential equations (23) in therms of account for the thermal conductivity and beam distribution in the hot spot.

## 6. Conclusions

1. Lifetimes of stripper targets under intensive nonstationary beams can be described by two failure mechanisms: radiation damage accumulation and evaporation of target. At the maximal temperatures less than 2500 $^oK$ the radiation damage are dominated; at temperatures above 2500 $^0K$ the mechanism of evaporation of a foil prevails.
2. The time structure of pulsed beam can influence on the lifetime.
3. The strength characteristics of carbon matirial show considerable scattering. However for the analysis of service life of stripper targets as it follows from expression (6), it is necessary to measure simultaneously the both values: the elastic module *M* and the ultimate strength $\sigma_P$.
4 For better discription of stripper targets behavior in therms of radiation damage it is important to take into account the foil microstructure, the conditions of fastening of a film on the film frame, the change of density of a material of a foil due to an irradiation.
5. It is necessary to carry our some experiments to check the dependence of the foil lifetime on $K_d$ and $\sigma_P$.

## References

1. N.R.S.Tait, D.W.L.Tolfree et al., Nucl. Instr. Meth., 167(1979) 21.
2. P. Maier-Komor, G. Dollinger, H.J.Korner, Nucl. Instr. Meth., A438 (1999 73.
3. I. Sugai et al, Nucl. Instr. Meth., A265 (1988 376.
4. G.Frick, Rev. Phys. Appl., 12(1977) 1525.
5. J. Yntema, F.Nickel, Lecture Notes in Physics, v.83, Experimental methods in heavy ion physics. Springer Verlag, Berlin, Heidelberg, New York, 1978.

6.  F.Nickel, Nucl. Instr. Meth., 195(1982) 457.
7.  J.Yntema, Nucl. Instr. Meth., 163(1979) 1.
8.  I.Sugai et al. Proc. of HB2006, Tsukuba, Japan.
9.  E.A.Koptelov, S.G.Lebedev, V.N.Panchenko, Nucl. Instr. Meth., A256 (1987 247.
10. E.A.Koptelov, S.G.Lebedev, V.N.Panchenko, Nucl. Instr. Meth., B42 (1989 239.
11. S.G.Lebedev, Nucl. Instr. Meth., B85 (1994 276.
12. S.G.Lebedev, Nucl. Instr. Meth., A362 (1995) 160.
13. S.G.Lebedev, Nucl. Instr. Meth., A397 (1997) 172.
14. F.F.Komarov, M.A.Kumakhov, Radiat. Eff., 22 (1974)1.
15. Korobenko VN and Savvatimskiy AI in Temperature: Its Measurement and Control in Science and Industry, vol. 7, edited by D. C. Ripple (American Institute of Physics Conference Proceedings, Melville, New York, 2003), pp. 783-788.
16. R. O. Dillon, Abbas Ali, N. J. Ianno, and A. Ahmad, J. Vac. Sci. Technol., A19 (2001) 2826.
17. S. Hoshino, K. Fujii, N. Shohata, H. Yuji, and Masahiro, J. Appl. Phys.65 (1989) 1918.
18. J. E. Field, in Diamond and Diamond-Like Films and Coatings NATO-ASI-Series, edited by R. E. Clausing, L. L. Horton, J. C.Angus, and P. Koidl, Plenum, New York, 1991, p.17.
19. Karl von Reden and Enid Sichel, Development of Carbon Nanotube Stripper Foils, Proc. of the Symposium of Northeastern Accelerator Personnel 2003, Strasbourg, FR.
20. T.Call, J.O.Stoner and S.Bashkin, Nucl. Instr. Meth., 167(1979) 33.
21. N.R.S.Tait, D.W.L.Tolfree and B.H.Armitage, Nucl. Instr. Meth., 163(1979) 1.
22. D.V.Fedoseev, S.P.Varnin, B.V.Deryagin, Fiziko-Himicheskaya Mechanika Materialov, 2(1977) 118 (in Russian).
23. F. Davanloo, T. J. Lee, D. R. Jander, H. Park, J. H. You, and C. B.Collins, J. Appl. Phys., 71(1992) 1446.
24. F. Rossi, B. Andre, A. Van Veen, P. E. Mijnarends, H. Schut, W. Gissler, J. Haupt, G. Lucazeau, and L. Abello, J. Appl. Phys., 75(1994) 3121.
25. C.J.Liaw, Y.Y.Lee and J.Tuozzolo, Proc. of Particle Accelerator Conference 2001, Chicago, IL, June 18-22, 2001, p. 1538.
26. B.P.Gikal, G.G.Gulbekyan, V.I.Kasacha, D.V.Kamanin, Preprint JINR P9-2005-110, Dubna, 2005 (in Russian).
27. C.J.Liaw, Y.Y.Lee, J.Alessi, J.Tuozzolo, Proc. Of Particle Accelerator Conference 1999, New York, p. 3300.

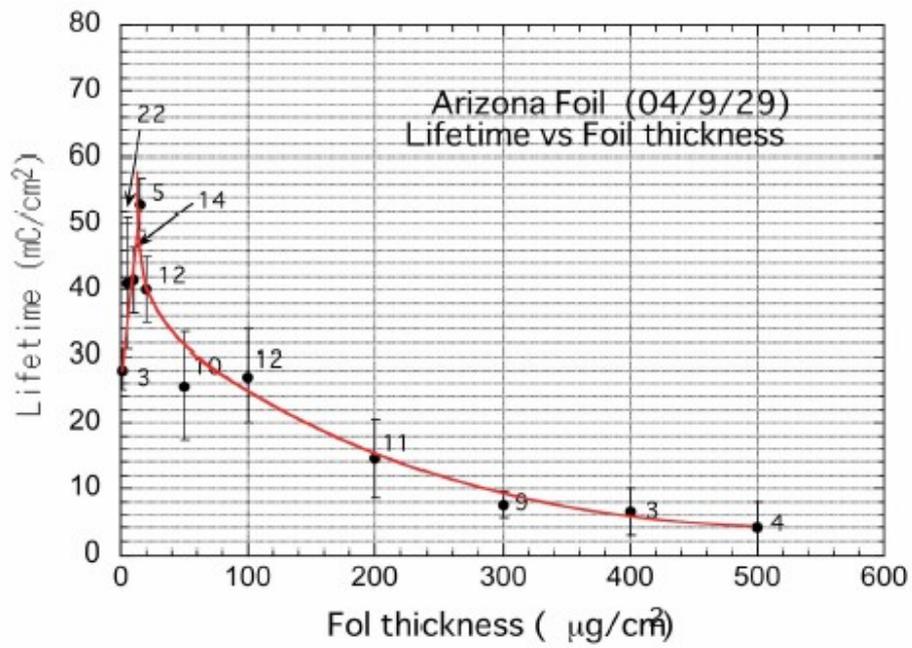

Fig.1. "Inverse" dependence [8] of lifetime vs. thickness of a target in the range of 25 -500 µg/cm$^2$, caused by transmission reduction of a particle beam.

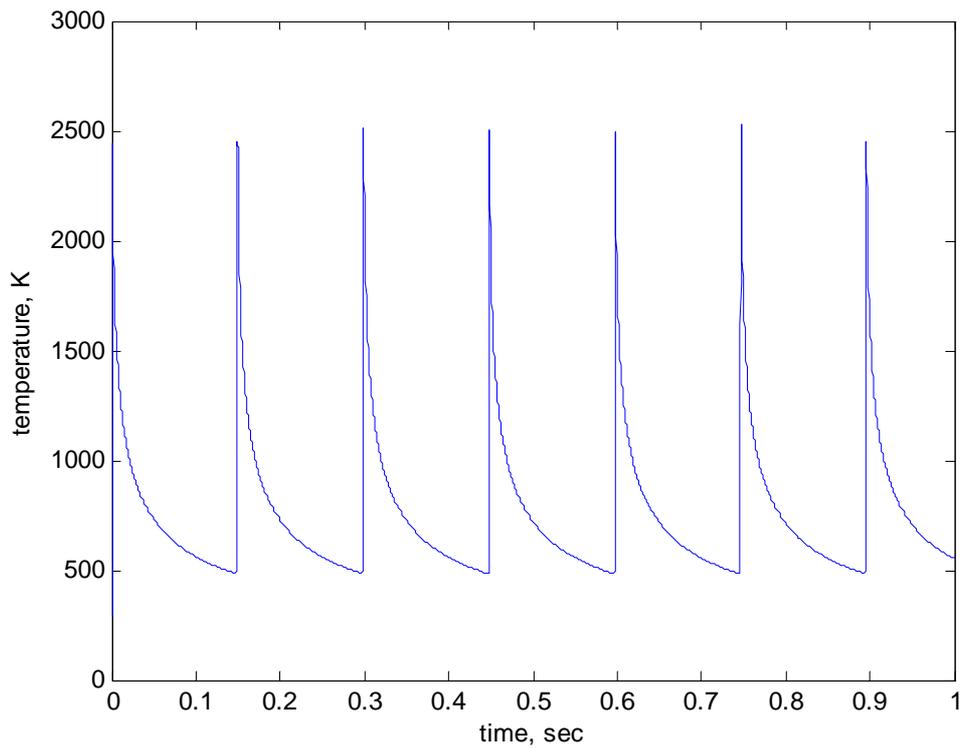

Fig.2. A temperature field of a *BNL* linac target in the first second of work at a pulse current 2 *mA*.

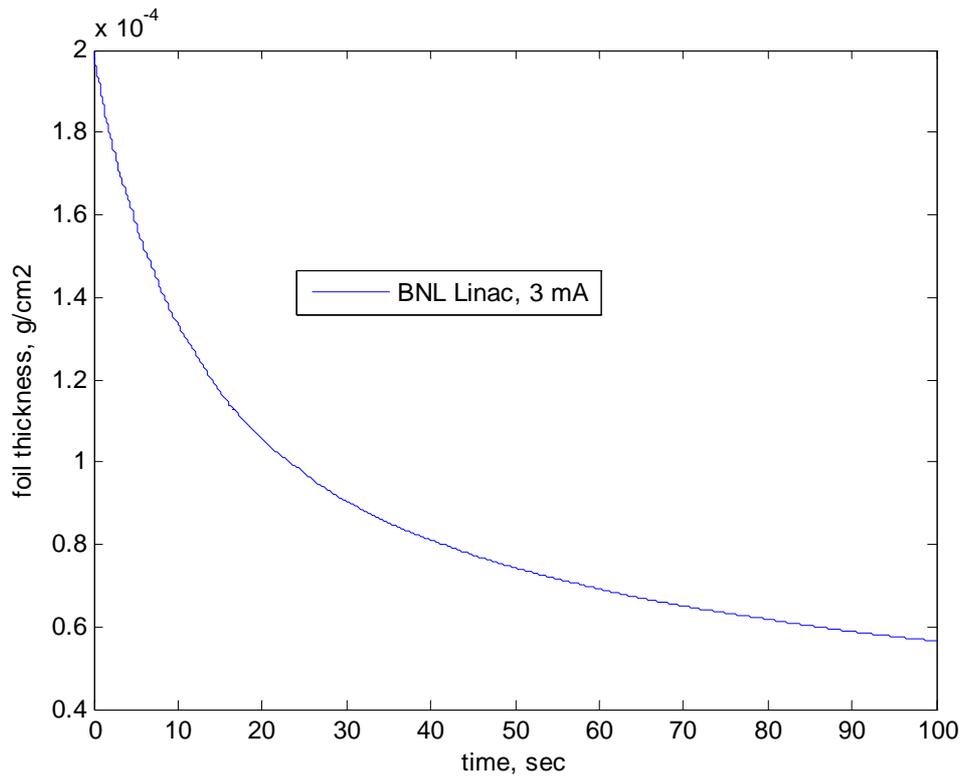

Fig.3. Nonlinear reduction of thickness of a foil at the BNL linac at a pulse current 3 *mA*, caused by the reduction of temperature under reduction of thickness.

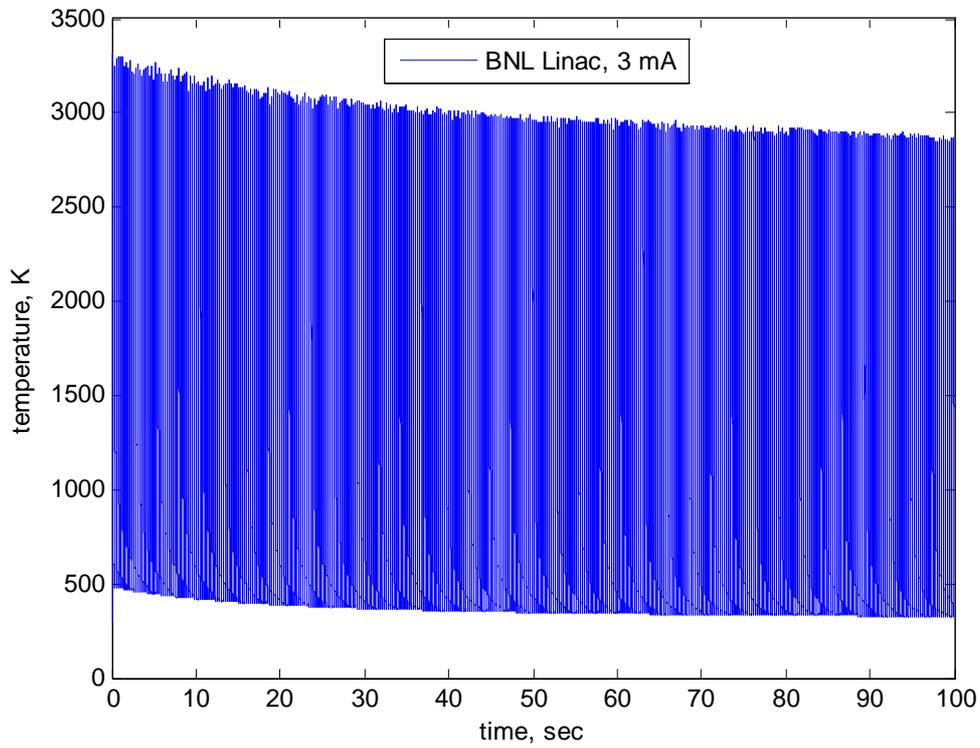

Fig.4. Deformation of a temperature field in a target of *BNL* linac, caused by the reduction of thickness of a foil due to its evaporation.

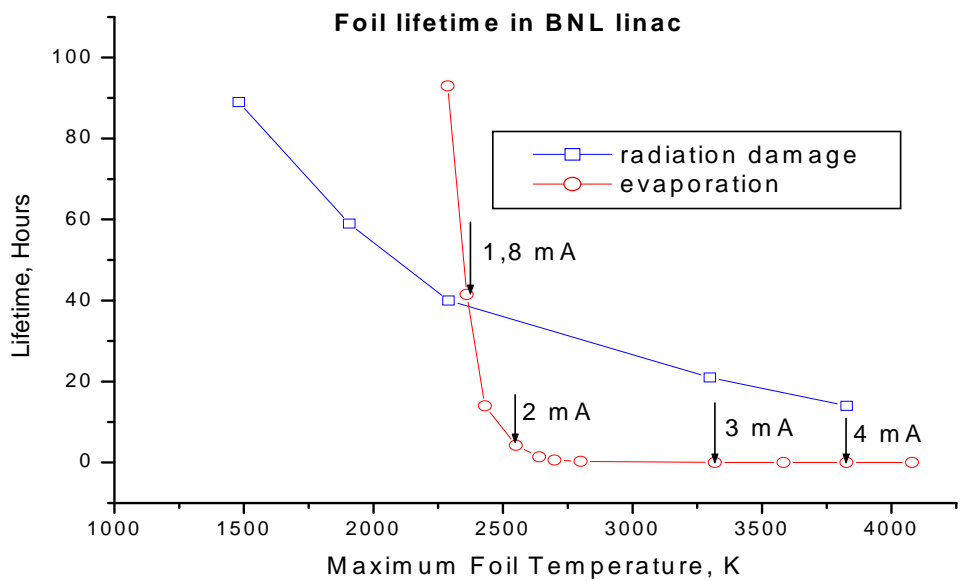

Fig.5. Calculated dependences of lifetime of *BNL* linac foil due to processes of radiation damage and evaporation.